\documentclass[twocolumn,prb]{revtex4-1}

\usepackage{graphicx,amssymb,soul,color,amsmath,bm}
\usepackage{epstopdf,hyperref}
\usepackage{braket,dsfont,nicefrac}
\usepackage[normalem]{ulem}
\usepackage[mathcal]{euscript}
\hypersetup{colorlinks=true,citecolor=blue,linkcolor=magenta}

\sethlcolor{yellow}
\allowdisplaybreaks

\usepackage{xcolor}

\newcommand{\kk}{\mathbf{k}}
\newcommand{\kr}{\mathbf{k}\cdot\mathbf{r}}
\newcommand{\vsigma}{\vec{\sigma}}

\begin{document}

\title{Chebyshev expansion of spectral functions using restricted Boltzmann machines}

\author{Douglas Hendry}
\affiliation{Department of Physics, Northeastern University, Boston, Massachusetts 02115, USA}

\author{Hongwei Chen}
\affiliation{Department of Physics, Northeastern University, Boston, Massachusetts 02115, USA}

\author{Phillip Weinberg}
\affiliation{Department of Physics, Northeastern University, Boston, Massachusetts 02115, USA}

\author{Adrian E. Feiguin}
\affiliation{Department of Physics, Northeastern University, Boston, Massachusetts 02115, USA}


\date{\today}
\begin{abstract}
Calculating the spectral function of two dimensional systems is arguably one of the most pressing challenges in modern computational condensed matter physics. While efficient techniques are available in lower dimensions, two dimensional systems present insurmountable hurdles, ranging from the sign problem in quantum Monte Carlo (MC), to the entanglement area law in tensor network based methods. We hereby present a variational approach based on a Chebyshev expansion of the spectral function and a neural network representation for the wave functions. The Chebyshev moments are obtained by recursively applying the Hamiltonian and projecting on the space of variational states using a modified natural gradient descent method. We compare this approach with a modified approximation of the spectral function which uses a Krylov subspace constructed from the ``Chebyshev wave-functions''.
We present results for the one-dimensional and two-dimensional Heisenberg model on the square lattice, and compare to those obtained by other methods in the literature. 
\end{abstract}

\maketitle

\section{Introduction}
The knowledge of the excitation spectrum of a quantum many-body system provides valuable information that be directly compared to experiments. While much effort is dedicated to the calculation of ground state wave functions, their energies do not have much experimental relevance since they cannot be measured.  Experiments such as photoemission or inelastic neutron scattering spectroscopies, for instance, measure energy differences between excited and ground states, a density of states. In addition, the integrated weight also yields the equal time correlations in the ground state, that can also indicate the existence of order or lack thereof, or the presence of quasiparticle excitations that could be used as fingerprints to identify quantum phases. Unfortunately, the numerical evaluation of dynamical correlation functions in strongly correlated systems is a very difficult undertaking, and the development of techniques that can applied beyond one dimension is possibly one of the pressing challenges in computational condensed matter. 

Since the advent of high-temperature supeconductivity, progress has been marked by ingenuity to overcome the computational limitations of the time. The first techniques to emerge were based on 
exact diagonalization\cite{Dagotto1994}, which is limited to small clusters, and different variants of quantum Monte Carlo, that suffers from the sign problem and requires uncontrolled analytic continuations \cite{Schuttler1986,Sandvik1998,Silver1990,Gubernatis1991,Syljuaasen2008,Fuchs2010,Sandvik2016,Shao2017}. Soon after the invention of the density matrix renormalization group (DMRG) \cite{White1992,White1993,Schollwock2005,Schollwock2011,Feiguin2013a}, several approaches to extract dynamical properties were proposed, that can be generically referred-to as dynamical DMRG\cite{Hallberg1995,Kuhner1999,Jeckelmann2002,Dargel2011,Dargel2012,Nocera2016}, or DDMRG. The time-dependent density matrix renormalization group and recent variations using Chebyshev expansions have been important developments, giving access to accurate spectra for very large one-dimensional systems \cite{Daley2004,White2004a,Feiguin2005,vietri,Feiguin2013b,Holzner2011,Wolf2015,Xie2018} and quasi-2D cylinders \cite{Zaletel2015,Verresen2018} with moderate computational resources. An alternative approach consists of  proposing variational forms for the excited states, which can be represented with matrix product states (MPS)\cite{Haegeman2013,Vanderstraeten2015,Vanderstraeten2015b,Vanderstraeten2019} or other variational wave functions that can be easily extended to higher dimensions and are free from the sign problem\cite{Li2010,DallaPiazza2014,Ferrari2018,Hendry2019}. The variational Monte Carlo (VMC) approach relies on a variational ansatz for the ground state inpired by some physical insight (typically of the form of a Gutzwiller projected wave function) and can provide a few hundred discrete poles by diagonalizing a Hamiltonian matrix projected onto a subspace of single magnon excitations.  

In this work, we bring together some of these ideas translated in the context of recent advances in quantum machine learning.  
Attempts in this direction have already demonstrated very promising results: In Ref.\onlinecite{Hendry2019}, the concepts behind correction-vector dynamical DMRG were generalized to arbitrary variational wave functions, and applied to the Heisenberg model using restricted Boltzmann machines (RBM) \cite{Carleo2017,Saito2017,Cai2018,Glasser2018}. 
RBMs are a type of artificial neural network widely used in machine learning that inspired 
Carleo and Troyer\cite{Carleo2017} to propose a variational wave-function for a spin-$\frac{1}{2}$ system of $N$ sites.  The visible layer corresponds to the spin configurations $\vsigma^z=(\sigma^z_1,\sigma^z_2,\cdots,\sigma^z_N)$. Then the coefficients of the wave function is given as $\psi(\vsigma^z,\vec{a},\vec{b},W) = e^{\sum_{i=1}^N a_i \sigma^z_i }\prod_{j=1}^M 2\cosh{(\theta_j)}$ where $\theta_j = b_j + \sum_{i=1}^N W_{ij}\sigma^z_i$. The weights $W_{ij}$  and the biases $A$ and $b$ in expression are free parameters that are used to variationally minimize the energy $E=\langle \psi|H\psi \rangle/\langle \psi|\psi\rangle$. This procedure is carried out using Monte Carlo to sample over all possible spin configurations. The remarkable aspect of this type of wave-functions is that they can encode a great deal of information in a relatively small set of variational parameters, without making any assumptions about the physics of the problem \cite{Chen2018}.

The main goal of this paper is to variationally approximate the zero temperature Green's function for a quantum many-body system:
\[
G_{ij}(z)=\langle\psi| \hat{A}^\dagger_j \frac{1}{z-\hat{H}} \hat{A}_i|\psi\rangle
\]
where $A_i$ and $A_j$ are some local operators of interest, $E_0$ is the ground state energy and $z=\omega+E_0+i\eta$. The spectral function is thus obtained as $A_{ij}(\omega) = -\frac{1}{\pi}G_{ij}(z)$. The correction vector DMRG method recasts this calculation as a complex system of equations for each frequency $\omega$. In Ref.\onlinecite{Hendry2019}, the solution to this system of equations was encoded in the form of an RBM. The optimization is carried out by means of a natural gradient descent approach based on quantum geometry concepts that allows one to solve a large system of equations stochastically with RBMs, or any other variational wave-function.  

We here explore a different avenue by expanding the spectral function in terms of Chebyshev polynomials. In Ref.\onlinecite{Holzner2011} it was shown that this approach, combined with DMRG, can yield very accurate results with a fraction of the effort compared to other methods. In fact, it comes with several advantages over other methods, such as: (i) it provides uniform resolution over the entire relevant frequency range with a relatively small numbers of Chebyshev moments and (ii) the information to calculate each moment is encoded in an RBM that can be stored for later use or systematically improved if needed. 

The paper is organized as follows: In Sec.\ref{sec:cheby} we introduce the basic ideas behind the Chebyshev representation of the spectral function; in Sec. \ref{sec:rbm} we discuss the implementation using RBMs, and the details about the variational optimization. Results can be improved by taking advantage of error correction measures described in Sec.\ref{sec:correction}. Sec. \ref{sec:corrvec} covers an alternative approach using a representation of the spectral function in terms of a Krylov expansion of the basis. Results for both the one-dimensional and two-dimensional Heisenberg model on the square lattice are presented in Sec.\ref{sec:results}. We finally close with a summary and discussion of possible strategies to improve the accuracy of the calculation. 

\section{Chebyshev expansion} \label{sec:cheby}

In this section we briefly describe the expansion of the spectral function $A_{ij}(\omega)$ in terms of Chebyshev moments. We refer the reader to Ref.~\onlinecite{Holzner2011} for a more detailed discussion. 
An important aspect of Chebyshev polynomials is they are defined in the interval $[-1,1]$. Hence, we rescale the frequencies and the Hamiltonian such that the range of the expansion encompasses the region with the bulk of the spectral weight. If our problem has non-zero spectral weight in the interval $0\le\omega \le W_A$, we define $\omega'=\omega/a-W'$ with $a=W_{*}/2W'$, and $H'=(H-E_0)/a-W'$. In these expressions, $W_{*} > W_A$ is an effective bandwidth larger than $W_A$, and $W'$ is the rescaled bandwidth, that we chose, for safety reasons, to be slightly smaller than 1 ($W'=1-0.0125$ in the following). 
In a nutshell, the spectral function can now be written as:
\begin{equation}
A_{ij}(\omega)=\frac{2W'/W_{*}}{\pi\sqrt{1-\omega'^2}}\left[g_0 \mu_0 + 2\sum_{n=1}^{N-1} g_n \mu_n T_n(\omega')\right]   
\label{Aw}
\end{equation}
where $T_n(x)$ is the $n^{th}$ Chebyshev polynomial of the first kind, the coefficients $g_n$ are damping factors that affect the broadening and $\mu_n$ are the corresponding Chebyshev moments:
\begin{equation}
    \mu_n = \langle \psi|\hat{A}^\dagger_j T_n(\hat{H}')\hat{A}_i|\psi\rangle.
\end{equation}
Luckily, one can exploit the definition of the Chebyshev polynomials and their recursion relation to recast the moments as:
\begin{equation}
    \mu_n=\langle \psi|\hat{A}_j|t_n\rangle, \,\,\, |t_n\rangle=T_n(\hat{H}')\hat{A}
    _i|\psi\rangle,
\end{equation}
with
\begin{eqnarray}
T_{n+1}(x) &=& 2xT_n(x) - T_{n-1}(x), \nonumber \\
T_0(x) &=& 1, T_1(x) = x.
\end{eqnarray}
Hence, given an initial wave-function $|t_0\rangle=\hat{A}_i|\psi\rangle$, we iteratively solve for a sequence of ``Chebyshev wave-functions'' $|t_1\rangle$,$|t_2\rangle$,... such that
$|t_n\rangle=T_{n}(\hat{H'})|t_0\rangle$. In the following section, we describe how to carry out this task using a variational representation of $|t_n\rangle.$

\section{Variational optimization}\label{sec:rbm}

Let  $|\phi\rangle $ be the target wave function for a variational wave function $|\psi(\alpha)\rangle$ with variational parameters $\{\alpha_1,\alpha_2\cdots \alpha_M\}\in\mathbb{C}^{M}$ that need to be optimized to make $|\psi(\alpha)\rangle$ as equal to $|\phi\rangle $ as possible up to an overall constant. This is done by minimizing Fubini-Study metric --essentially the angle-- between the two wave functions, which is given by

$$\gamma(\psi,\phi)=\arccos{ \sqrt{ \frac{\langle\psi|\phi\rangle\langle\phi|\psi\rangle}{ \langle\psi|\psi\rangle\langle\phi|\phi\rangle }} }.$$
Taking the gradient of ${\gamma}^2$ with respect $|\psi\rangle$ gives 

$$|\delta\psi\rangle=\delta\gamma \cdot \,\gamma(\psi,\phi)\cot{(\gamma(\psi,\phi))}\left[\frac{\langle\psi|\psi\rangle}{\langle\psi|\phi\rangle}|\phi\rangle-|\psi\rangle\right],
$$
which is a differential rotation by an angle $\delta\gamma \cdot \gamma$ towards  $|\phi\rangle$ (with a change in the overall phase to match $|\psi\rangle$). As with rotations in real space, $\langle\psi|\delta\psi\rangle=0$ and $\langle\delta\psi|\delta\psi\rangle={(\delta\gamma\cdot\gamma)}^2 \langle\psi|\psi\rangle$. 

The procedure to update the variational coefficients $\alpha$ using natural gradient descent is described in detail in Ref.~\onlinecite{Hendry2019} and we hereby summarize it. The updates $\Delta\alpha$ correspond to projecting $|\delta\psi\rangle$ onto the manifold of possible variational wave-functions of the proposed form (in our case, an RBM). To calculate $\Delta\alpha$ one must solve the following linear system of equations:
$$ 
\sum_{k'=1}^{M} S_{kk'}\Delta\alpha_{k'}=
\frac{\langle \partial_k \psi|\delta\psi\rangle}{\langle\psi|\psi\rangle} -\frac{\langle \partial_k \psi|\psi\rangle}{\langle\psi|\psi\rangle}\frac{\langle  \psi|\delta\psi\rangle}{\langle\psi|\psi\rangle} , 
$$
where $|\partial_k\psi\rangle=\frac{\partial}{\partial\alpha_k}|\psi\rangle$ and $S$ is given by 
$$ S_{kk'} = \frac{\langle \partial_k \psi| \partial_{k'} \psi \rangle}{\langle\psi|\psi\rangle} -\frac{\langle \partial_k \psi|\psi \rangle}{\langle\psi|\psi\rangle} \frac{\langle \psi| \partial_{k'} \psi \rangle}{\langle\psi|\psi\rangle}.
$$

\begin{figure}
    \centering
    \includegraphics[width=0.48\textwidth]{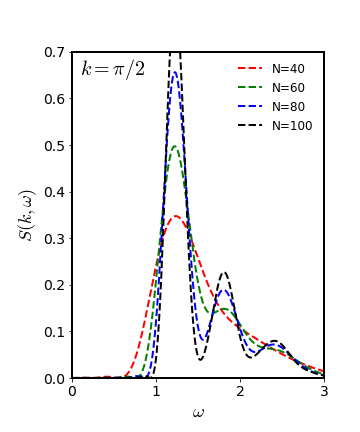}
    \caption{Evolution of the spin dynamic structure factor of a Heisenberg chain with the number of Chebyshev moments using RBM wave-functions for $L=32$ at momentum $k=\pi/2$.}
    \label{fig:fig1}
\end{figure}
To estimate the the overlap between wavefunctions in the definition of $S$ and in the linear system defined above, we use VMC. The states $|s\rangle$ are scampled according $P(s)={|\psi_s|}^2 / \langle\psi|\psi\rangle$, where $\psi_s \equiv \langle s|\psi\rangle$.  Using these sampled states we define estimators such as the log derivative $\mathcal{O}_k(s)=\frac{\partial\psi_s}{\partial\alpha_k} / \psi_s$ and the ratio $R(s)=\phi_{s} / \psi_{s}$ that are used to calculate the following relevant quantities:
\begin{eqnarray}
S_{kk'} &=& \langle\mathcal{O}_k(s)^{*}\mathcal{O}_{k'}(s)\rangle-\langle \mathcal{O}_k(s)\rangle^{*}\langle\mathcal{O}_{k'}(s)\rangle, \nonumber \\
\gamma(\psi,\phi) &=& \arccos{ \sqrt{\frac{{|\langle R(s)\rangle|}^2}{\langle{|R(s)|}^2\rangle}}},\\
\frac{\langle \partial_k \psi|\delta\psi\rangle}{\langle\psi|\psi\rangle} &=& \delta\gamma \,\gamma(\psi,\phi)\cot{(\gamma(\psi,\phi))}\times\nonumber\\ &&
\left[\frac{\langle \mathcal{O}_k(s)^{*}R(s)\rangle}{\langle R(s) \rangle}-\langle \mathcal{O}_k(s)^{*} \rangle \right]. \nonumber
\end{eqnarray}

Notice that when $|\psi\rangle\propto |\phi\rangle$, the variance of $R(s)$ goes to $0$. As a consequence, the variance of the estimators above (except $S$) also goes to zero. Once the parameters have converged, the normalization constant $\beta$ can be calculated as
$$ \beta=\frac{\langle\psi|\phi\rangle}{\langle\psi|\psi\rangle}=\langle R(s) \rangle.$$

This optimization algorithm is applied to every step of the Chebyshev recursion: Each wave-function in sequence is approximated by an non-normalized variational wave function $|\tilde{t}_n\rangle$ and corresponding normalization constant $\beta_n$.  The wave-functions are solved iteratively using the Chebyshev recurrence relation and is projected onto an RBM form such that the corresponding target wave functions are $|t_1\rangle=\beta_{0}{H'}|\tilde{t}_0\rangle$, and  $|t_n\rangle=2\beta_{n-1}{H'}|\tilde{t}_{n-1}\rangle-\beta_{n-2}|\tilde{t}_{n-2}\rangle$ for $n\geq2$, where $\beta_0=1$ and $|\tilde{t}_0\rangle=|t_0\rangle$. These wave-functions can be easily stored, since the number of variational coefficients is $\mathcal{O}(N^2)$. This allows the calculation to proceed in three steps: First, the wave functions $|t_n\rangle$ are obtained; second, the Chebyshev moments are calculated by sampling over the wave functions to obtain the overlaps and, finally, the spectral function can be reconstructed.


\section{Chebyshev Moment Correction} \label{sec:correction}
In the absence of errors, only the overlap of the Chebyshev vectors with $|t_0\rangle$ is needed to obtain the Chebyshev moments. Due to limitations in the expressivity of the RBM wave-function and the fact that each successive Chebyshev wave-function is obtained recursively, the variational Chebyshev wave-functions will become a less accurate approximation, amplifying the error for the higher moments. However, we can obtain better estimates for the Chebyshev moments by recalculating the Chebyshev wave-functions in the sub-space spanned by the original ones. Thus the errors are no longer dependent on the individual accuracy of each variational Chebyshev wave-function, but will depend on how well the true Chebyshev wave-functions can be represented in the new basis.


The procedure to recalculate the Chebyshev moments is as follows: The projector for the space spanned by the RBMs is given by $\hat{P} = \sum_{mn} (\sigma^{-1})_{mn} |t_m \rangle \langle t_{n}|$, $\sigma_{nm}=\langle t_n|t_m\rangle$ and $H_{nm}=\langle t_n|\hat{H}|t_m\rangle$.  Let us project the Hamiltonian onto this basis as $\hat{H}_P=\hat{P}\hat{H}\hat{P}$. Then, the recalculated Chebyshev wave-functions are given by 
$$ |t'_n \rangle = T_{n}(\hat{H}_P)|t'_0 \rangle=\sum_{m} c_{mn}|t_m \rangle $$
with $|t'_0 \rangle = |t_0 \rangle$ and thus $c_{0n}=\delta_{0n}$.  The coeficients $c_{mn}$ can be obtained from the Chebyshev recursion relations which results in the following system of equations
\begin{gather}
    \sum_{l} \sigma_{ml}c_{l1} = \sum_{l} H_{ml}c_{l0}\\
    \sum_{l} \sigma_{ml}c_{l,n+1} = \sum_{l} 2H_{ml}c_{ln}-\sigma_{ml}c_{l,n-1}
\end{gather}
Finally the moments are obtained as 
$$\mu_{n} =  \sum_{ml} c_{m0}^{*}\sigma_{ml}c_{ln}.$$

\section{Continued fraction}\label{sec:corrvec}

Another avenue we can explore in order to approximate the spectral function from the Chebyshev wave-functions is the vector correction method\cite{Kuehner1999,Jeckelmann2002,Hendry2019}. The vector correction method involves finding the wave function that is the solution to the system of equations
$(z-\hat{H})|\chi_j(z)\rangle = \hat{A}_j |\psi_{gs}\rangle$.  Then, the Green's function is given by the overlap $$
G_{ij}(z)=\langle \psi_{gs}|\hat{A}_i^{\dagger}|\chi_j(z) \rangle
$$

Since Chebyshev wave-functions form a non-orthogonal basis for the Krylov subspace, we can obtain an approximate solution for $|\chi_j(z)\rangle$ by utilizing a Krylov expansion. First, the Lanczos wave functions $|v_0\rangle,\dots ,|v_{N-1}\rangle$ are obtained from the projected Hamiltonian $\hat{H}_{P}$ with $|v_0\rangle=|t_0\rangle$, giving the matrix elements of the Hamiltonian in tridiagonal form.
Finally, to obtain the spectral function from $H_P$ we use a continued-fraction expansion of the spectral function, which is equivalent to solving the system of equations projected onto this subspace\cite{Dagotto1994,Dargel2012}.



\section{Results} \label{sec:results}

\begin{figure}
    \centering
    \includegraphics[width=0.48\textwidth,trim=80 50 50 50,clip]{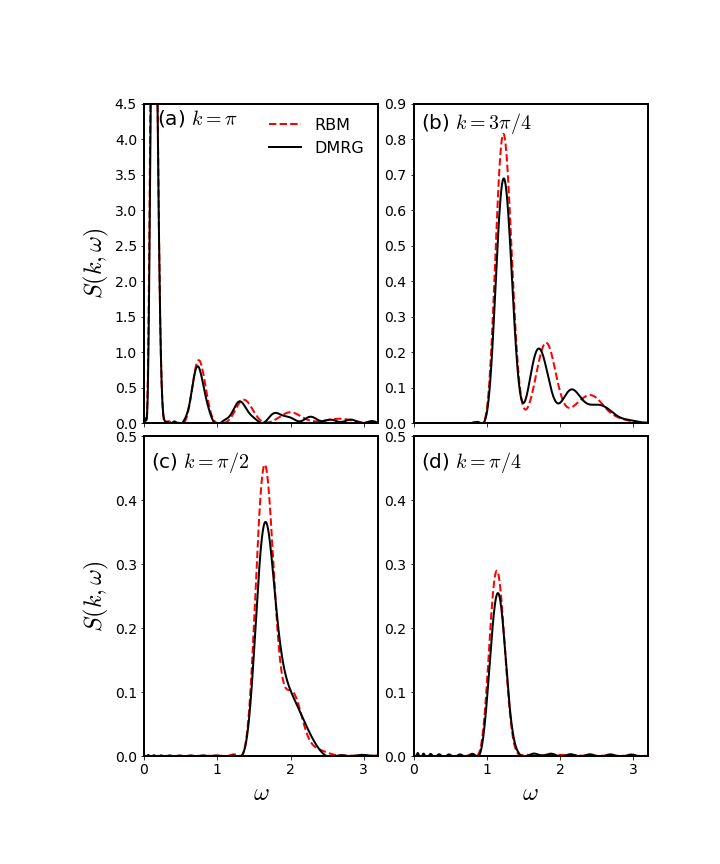}
    \caption{Spin dynamic structure factor for a Heisenberg chain obtained using $N=100$ RBM wave functions, compared to DMRG results using a Chebyshev expansion for $L=32$ and various values of momenta $k$. }
    \label{fig:fig2}
\end{figure}

\begin{figure}
    \centering
    \includegraphics[width=0.48\textwidth,trim=80 50 50 50,clip]{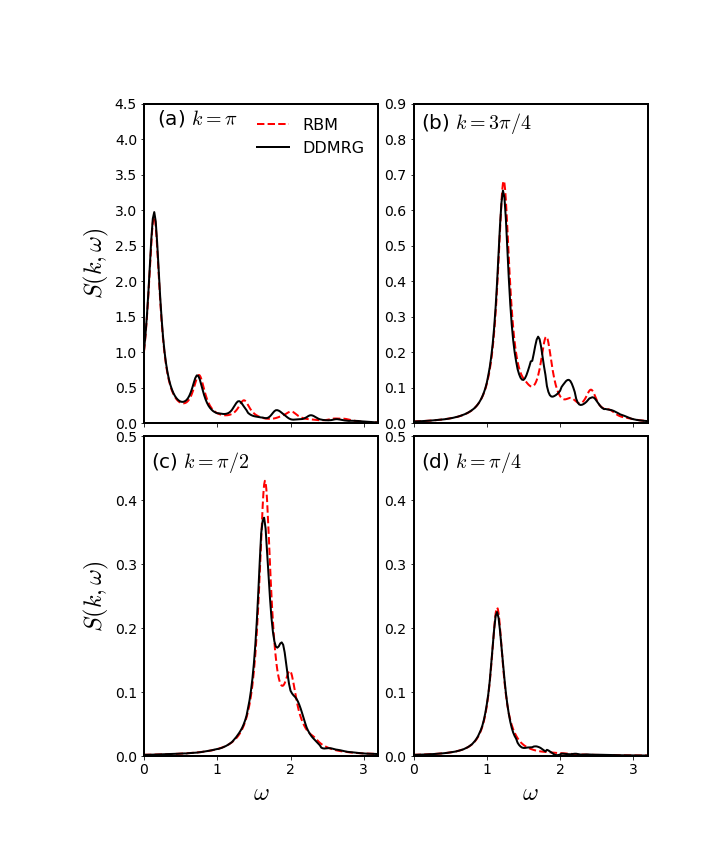}
    \caption{Spin dynamic structure factor for a Heisenberg chain obtained with a continued fraction approach in the basis of $N=100$ Chebyshev RBM wave-functions, compared to dynamical (correction vector) DMRG, for $L=32$ and various values of momenta $k$. An artificial broadening $\eta=0.1$ is introduced. }
    \label{fig:fig3}
\end{figure}

\begin{figure}
    \centering
    \includegraphics[width=0.48\textwidth,trim=80 50 50 50,clip]{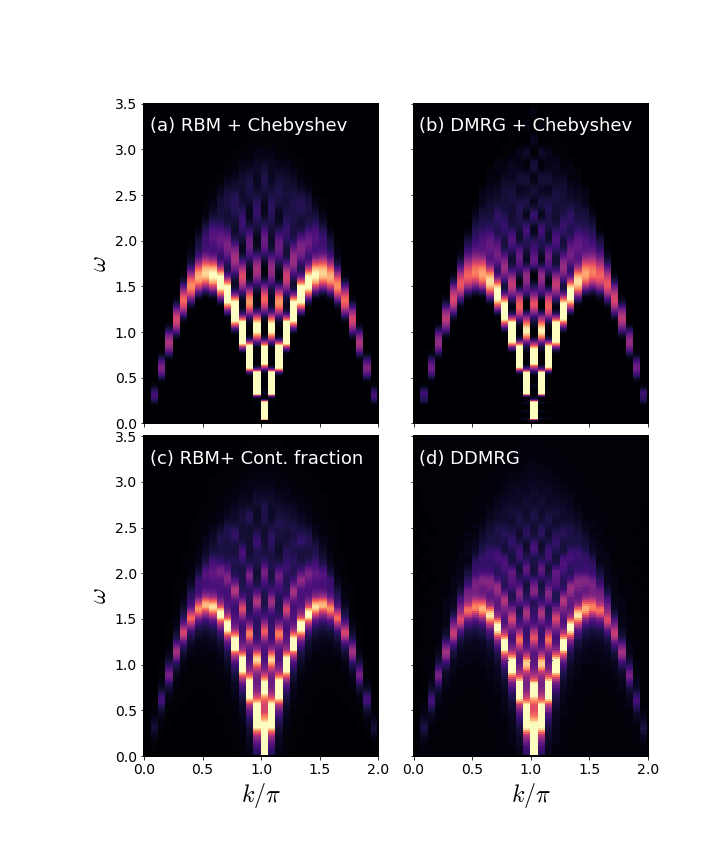}
    \caption{Color density plots depicting the spin dynamic structure factor for a Heisenberg chain with $L=32$ as a function of momentum and frequency, obtained by the four methods used in this work: (a) Chebyshev moments with RBM wave functions, (b) Chebyshev moments using DMRG, (c) continued fraction using the Chebyshev RBM basis and (d) dynamical DMRG. We use $N=100$ Chebyshev moments. }
    \label{fig:fig4}
\end{figure}

\begin{figure}
    \centering
    \includegraphics[width=0.48\textwidth,trim=80 50 50 50,clip]{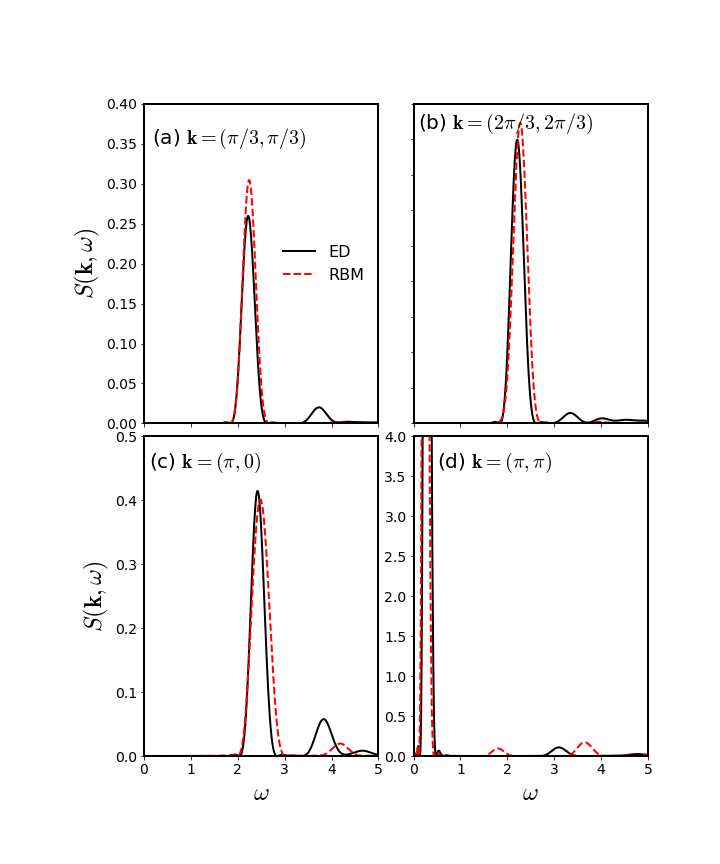}
    \caption{Spin dynamic structure factor for the Heisenberg model on a $L_x \times L_y = 6 \times 6$ square lattice using a Chebyshev expansion and exact diagonalization. }
    \label{fig:fig5}
\end{figure}

\begin{figure}
    \centering
    \includegraphics[width=0.48\textwidth,trim=80 50 50 50,clip]{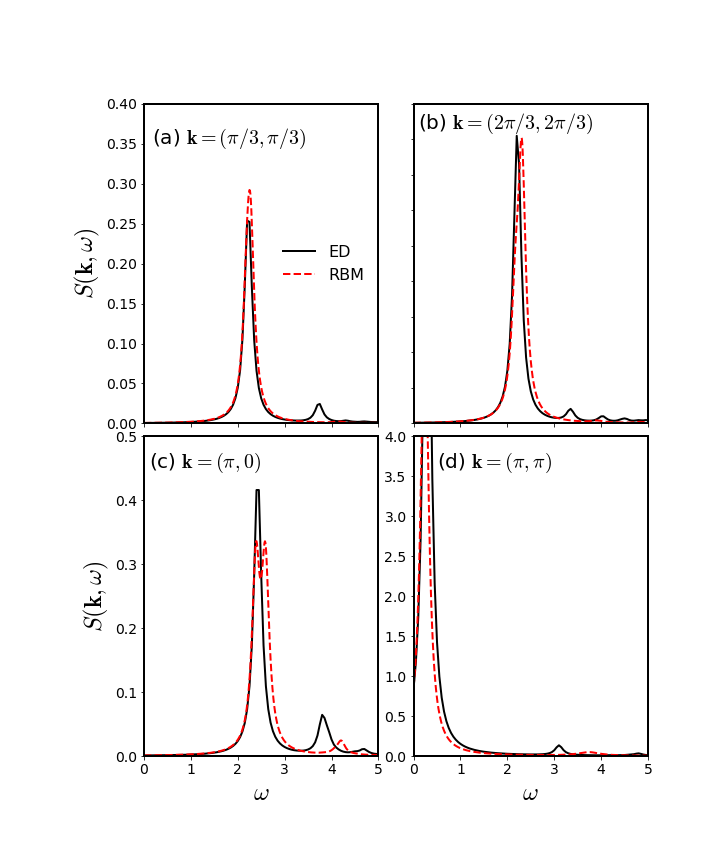}
    \caption{Spin dynamic structure factor for the Heisenberg model on a $L_x \times L_y = 6 \times 6$ square lattice using both continued fraction with RBM Chebyshev wave functions and exact diagonalization. }
    \label{fig:fig6}
\end{figure}

For demonstration and benchmarking we focus on the  spin-$\frac{1}{2}$ Heisenberg model in one and two dimensions on the square lattice:
\begin{equation}
\hat{H}=J\sum_{\langle ij\rangle} \vec{S}_{i}\cdot\vec{S}_{j},
\end{equation}
where $\vec{S}=(\hat{S}^x,\hat{S}^y,\hat{S}^z)$ are spin operators and the sum runs over pairs of nearest neighbor sites on the lattice. We consider periodic boundary conditions and chose $J$ as our unit of energy.
We are interested in the longitudinal spin structure factor, defined as:
\[
S^z(\kk,\omega) = -\frac{1}{\pi N} \mathrm{Im} \sum_n e^{i\kr_n}{\langle \psi| \hat{S}^z_0 \frac{1}{z-\hat{H}} \hat{S}^z_n |\psi \rangle}.
\]
where we have used translational invariance.
In our calculations, unless otherwise stated, we consider $W'=1-\epsilon/4$ with $\epsilon=0.05$; $W_*=10W'$. In addition, we use the so-called ``Jackson damping'' in Eq.(\ref{Aw}): 
$$g_n = \frac{(N - n + 1) \cos{\frac{\pi n}{N+1}} + \sin{\frac{\pi n}{N+1}} \cot{\frac{\pi}{N+1}}}{ N + 1},$$ 
where  $N$ is the number of moments used. We also implement spatial symmetries to improve the accuracy of our results, as described in Appendix \ref{app:symmetries}.

In Fig.\ref{fig:fig1} we show the evolution of the longitudinal structure factor of a 1D spin chain of length $L=32$ for a choice of momentum $k=\pi/2$ as a function of the number of Chebyshev moments $N$.
We clearly observe that the spectrum develops oscillations reproducing the well resolved high energy features in the finite system. These oscillations become more pronounced when increasing $n$ over the range of energies corresponding to the spinon continuum. This shows that one can obtain very reasonable results over the entire frequency domain with moderate effort using a small number of wave-functions, or Chebyshev iterations.

In Fig.\ref{fig:fig2} we compare results with $N=100$ to those obtained using the same approach but with DMRG as a solver ({\it i.e.}, using matrix product states as a variational ansatz). The agreement over the entire range of momenta and frequencies is remarkable. Notice that there is no straight forward
way to quantify the error, except by comparing to such accurate calculations using other methods as benchmark (DMRG in this case). We discuss the possible sources of the discrepancies in the discussion, Sec.\ref{sec:conclusion}.

Fig.\ref{fig:fig3} shows the same spectra but using the continued fraction approach, compared to dynamical DMRG. We note that while the positions of the peaks are the same, the profiles are different to those in Fig.\ref{fig:fig2} because the spectrum in this case is described by a superposition of Lorentzians. For all practical purposes, both techniques reproduce the same data. This is more dramatically illustrated in Fig.\ref{fig:fig4} where we compare all approaches as a function of momentum and frequency in color density maps. It is fair to say that they are practically indistinguishable. 

While results in 1D serve as a benchmark and illustrate the accuracy of the approach, we aim at making this method also applicable in higher dimensions, where one can access richer physics, such as quantum spin liquids\cite{Szabo2020}. In this case, we focus on a $6 \times 6$ square lattice, where our variational results can be contrasted directly to exact diagonalization (see Appendix ~\ref{app:ED}). In Figs.\ref{fig:fig5} and \ref{fig:fig6} we display a comparison for several values of momenta on the two-dimensional Brillouin zone using both, a Chebyshev expansion and a continued fraction approach, respectively. We find that while the first peak is very accurately described, the high energy features with smaller spectral weight are partially lost.

\section{Summary and conclusions}\label{sec:conclusion}

We have presented a highly efficient approach to calculate spectral functions of strongly correlated systems using a Chebyshev expansion of the Green's function and a variational representation of the many-body states in the form of restricted Boltzmann machines. Unlike a previous method introduced in the literature by the authors\cite{Hendry2019}, these calculations are numerically inexpensive and yield very accurate results in 1D. We have also described remedial measures to improve the accuracy of the results, such as using spatial symmetries, and correcting for the limited ``expressivity'' of the neural network.  
As mentioned earlier, the method provides uniform resolution over the entire relevant frequency range with a relatively small numbers of Chebyshev moments and the information to calculate each moment is encoded in an RBM that can be stored for post-processing.

In order to understand the sources of error we start by recalling that (i) our calculations rely on a particular form of the variational wave-function, meaning that the results will depend on the representation power of a RBM; (ii) the optimization of the wave function depends on our ability to accurately minimize our loss function with respect of the variational parameters. This loss function may have a complex landscape with local minima and, in addition, (iii) we need to consider the errors introduced by the stochastic sampling of all the different quantities. Due to the recursive nature of the calculation, since every wave function depends on the previous ones, any errors are propagated down the sequence. We have shown that these errors can be mitigated by recalculating the moments in a Chebyshev basis that has support outside the space of single RBMs. 

These ideas are general, as demonstrated by both prior studies using matrix product states and in this work using RBMs. The main challenge moving forward is to scale these methods to larger systems, which will be primarily determined by our ability to train larger networks, or variational wave functions, with a growing number of free parameters. 

It also worth noting that one can take existing moments and approximate higher moments using extrapolation techniques~\cite{Ganahl2014,Wolf2015}. This can help push calculations to larger system sizes by reducing the total number of moments required. 

We finally point out that the overall technique, including the error correction approach we used to improve the Chebyshev wave functions, is generic and can be implemented with any variational form of  wave-function.



\acknowledgements

AEF and HC acknowledge the National Science Foundation for support under grant No. DMR-1807814. DH is supported by a Northeastern Tier 1 grant.

\appendix
\section{Implementing symmetries}\label{app:symmetries}

The RBM wave functions are optimized to approximate $ |t_{n00} \rangle = T_n(\hat{H}) S^z_{00} |\psi_{gs} \rangle$ for the $n^{th}$ Chebyshev polynomial $T_n$ and approximate ground state $|\psi_{gs} \rangle$. To obtain the Chebyshev wave functions for momentum $\kk = (k_1,k_2)$,  we make use of the translational symmetry of the problem by assuming the ground state has zero momentum (translations do no result in a change in overall phase). Here we use $\hat{T}_{m_1m_2}$ to represent the translation operator on the square lattice with a shift of $(m_1,m_2)$, while $T_{n}(\hat{H})$ denotes the Chebyshev polynomials of the Hamiltonian. Using this notation we can write: 
\begin{widetext}
$$ |\widetilde{t}_{nk_1k_2} \rangle = \frac{1}{L} \sum_{m_1,m_2=0}^{L-1} e^{-i \frac{2\pi}{L} (k_1 m_1 + k_2 m_2 ) } \hat{T}_{m_1 m_2} \hat{S}^{z}_{00} \hat{T}_{-m_1 -m_2} T_n(\hat{H})\hat{S}^{z}_{00} |\psi_{gs}\rangle  = L \hat{P}^{(T)}_{k_1k_2} |t_{n00}\rangle $$
\end{widetext}
In this equation one can see that we are acting with the projection operator onto the momentum sector with momentum quantum numbers given by integers $k_1,k_2$ defined as:
$$ P^{(T)}_{k_1k_2} = \frac{1}{L^2} \sum_{m_1,m_2 =0}^{L-1} e^{-i \frac{2\pi}{L} (k_1 m_1 + k_2 m_2 ) } \hat{T}_{m_1 m_2}. $$
Thus, each momentum wave-function can represented in terms of linear combinations of translations of a single RBM optimized for the local $S^z_{00}$ operator.  Additionally, since $\hat{P}^{(T)}_{k_1k_2}$ is a projector, the resulting state will have the correct momentum and all the error from different momentum sectors will be projected out.  In the same vein, rotational symmetry can be enforced by applying the following projector of rotations and translations. 

$$ \hat{P}^{(R)} = \frac{1}{4} (1+\hat{T}_{10}\hat{R} + \hat{T}_{11}\hat{R}^2 + \hat{T}_{01}\hat{R}^3). $$
This projector is an average of each rotation followed by the translation which takes the $S^z$ operator back to the origin.  In the exact calculation for the Chebyshev wave-functions, the application of $\hat{P}^{(R)}$ should leave the $|t_{n00} \rangle$ unchanged. However, the RBM representation will not necessarily have rotational symmetry, therefore, the projector will correct any errors related to an absence of rotational symmetry.  After including rotation projector we find:
$$ |\widetilde{t}_{nk_1k_2} \rangle = L \hat{P}^{(T)}_{k_1k_2} \hat{P}^{(R)} |t_{n00}\rangle$$

All that is needed to calculate the Chebyshev moments is the overlap:
$$ \mu_{nk_1k_2} = \frac{\langle\widetilde{t}_{0k_1k_2} |\widetilde{t}_{nk_1k_2} \rangle}{\langle \psi_{gs}|\psi_{gs} \rangle} $$
However, due the recursive process of obtaining the Chebyshev wave-functions, the error will propagate down the sequence of wave-functions.  As mentioned in the main text, we can correct some of this error by constructing new Chebyshev wave-functions as linear combinations of the original RBM wave-functions. This error correction requires us to calculate the following quantities: 
$$ G^{(k_1k_2)}_{nn'}=\frac{\langle\widetilde{t}_{nk_1k_2} |\widetilde{t}_{n'k_1k_2} \rangle}{\langle \psi_{gs}|\psi_{gs} \rangle} $$
$$ H^{(k_1k_2)}_{nn'}=\frac{\langle\widetilde{t}_{nk_1k_2} |{H}|\widetilde{t}_{n'k_1k_2} \rangle}{\langle \psi_{gs}|\psi_{gs} \rangle} $$
These matrices are calculated by sampling over spin configurations states $|s\rangle$ probability $P(s)=\frac{\langle\psi_{gs}|s\rangle\langle s|\psi_{gs}\rangle}{\langle \psi_{gs}|\psi_{gs} \rangle}$ . Then, for each translation $(m_1,m_2)$, and rotation $m_r$ of $s$ we obtain:
$$B_{nm_1m_2m_r}(s) = \frac{\langle s|\hat{T}_{m_1m_2}\hat{R}^{m_r}|t_{n00}\rangle}{\langle s|\psi_{gs}\rangle}  $$
From these values, the ratio $\widetilde{B}_{nk_1k_2}(s)=\langle s|\widetilde{t}_{nk_1k_2} \rangle / \langle s|\psi_{gs}\rangle$ can be obtained by Fourier transforming $B$.
\begin{widetext}
$$
\widetilde{B}_{nk_1k_2}(s) = \frac{1}{L} \sum_{m_1,m_2 =0}^{L-1} e^{-i \frac{2\pi}{L} (k_1 m_1 + k_2 m_2 ) }\frac{1}{4}
\left[
B_{n,m_1,m_2,0}(s)+\right. \\ \left.B_{n,m_1+1,m_2 ,1}(s)+B_{n,m_1+1,m_2+1,2}(s)+B_{n,m_1,m_2+1,3}(s)
\right]
$$
\end{widetext}

%

Finally, the average gives $G^{(k_1k_2)}_{nn'}= \langle \widetilde{B}_{nk_1k_2}(s)^* \widetilde{B}_{n'k_1k_2}(s)\rangle$. A similar procedure is followed to obtain $ H^{(k_1k_2)}_{nn'}$.

\section{Exact Calculation for 2D Heisenberg Model}\label{app:ED}

In order to calculate the spectral function for $6\times 6$ lattice with periodic boundary conditions we implement the translational and point-group symmetries to block diagonalize the Hamiltonian using the QuSpin exact diagonalization library~\cite{quspin1,quspin2}. Any local operator $A_i$ explicitly breaks translational invariance, so instead we work with linear combinations of the local operators:
\begin{equation}
    \hat{A}_{k_1,k_2} = \frac{1}{L}\sum_{n_1,n_2=0}^{L-1} e^{-i \frac{2\pi}{L}\left(k_1 n_1 +k_2 n_2\right)}\hat{A}_{n_1,n_2}.
\end{equation}
The resulting operators have a well defined quantum number for the momentum/point-group symmetries, therefore, when applying this operator to the ground state, the resulting state lives in the sector with the momentum given by the operator:
\begin{equation}
    |A(k_1,k_2)\rangle = \hat{A}_{k_1,k_2}|\psi_{gs}\rangle
\end{equation}
With this state calculated we can calculate the Chebyshev moments as well as solve for the exact spectral function by using an iterative linear solver on the following linear equation:
\begin{equation}
    (z-\hat{H})|A(k_1,k_2)\rangle = |B(k_1,k_2,z)\rangle
\end{equation}
which can then be used to calculate the spectral function defined in momentum space:
\begin{multline}
    G_{k_1,k_2}(z) = \frac{1}{\pi} \langle\psi_{gs}|\hat{A}^\dagger_{k_1,k_2}\frac{1}{z-\hat{H}}\hat{A}_{k_1,k_2}|\psi_{gs}\rangle\\=\frac{1}{\pi}\langle A(k_1,k_2)|B(k_1,k_2,z)\rangle
\end{multline}


%
\end{document}